\documentclass{osa-article}

\journal{optica}
\articletype{Research Article}
\begin{document}

\title{Arbitrary cylindrical vector beam generation enabled by polarization-selective Gouy phase shifter}

\author{Junliang Jia,\authormark{1,2,4} Kepeng Zhang,\authormark{1,4} Guangwei Hu,\authormark{2} Maping Hu,\authormark{1,4}  Tong Tong,\authormark{1} Quanquan Mu,\authormark{3}  Hong Gao,\authormark{4} Fuli Li,\authormark{4} Cheng-Wei Qiu\authormark{2,**} and Pei Zhang\authormark{1,4,*}}

\address{
\authormark{1} MOE Key Laboratory for Nonequilibrium Synthesis and Modulation of Condensed Matter, School of Physics, Xi’an Jiaotong University,
Xi’an 710049, China\\
\authormark{2} Department of Electrical and Computer Engineering, National University of Singapore, Singapore, Singapore\\
\authormark{3} State Key Laboratory of Applied Optics, Changchun Institute of Optics, Fine Mechanics, and Physics, Chinese Academy of Sciences, Changchun, Jilin 130033, China\\
\authormark{4} Shaanxi Key Laboratory of Quantum Information and Quantum Optoelectronic Devices, School of Physics, Xi'an Jiaotong University, Xi'an 710049, China\\
}

\email{\authormark{*}zhangpei@mail.ustc.edu.cn\\ %% email address is required
\authormark{**}chengwei.qiu@nus.edu.sg} %% email address is required


\begin{abstract}
Cylindrical vector beams (CVBs), which possesses polarization distribution of rotational symmetry on the transverse plane, can be developed in many optical technologies. 
Conventional methods to generate CVBs contain redundant interferometers or need to switch among diverse elements, thus being inconvenient in applications containing multiple CVBs. 
Here we provide a passive polarization-selective device to substitute interferometers and simplify generation setup. 
It is accomplished by reversing topological charges of orbital angular momentum based on  polarization-selective Gouy phase.
In the process, tunable input light is the only condition to generate CVB with arbitrary topological charges. 
To cover both azimuthal and radial parameters of CVBs, we express the mapping between scalar Laguerre-Gaussian light on basic Poincar\'e sphere and CVB on high-order Poincar\'e sphere. 
The proposed device simplifies the generation of CVBs enormously, and thus has potentials in integrated devices for both quantum and classic optical experiments. 
\end{abstract}

%%%%%%%%%%%%%%%%%%%%%%%%%%  body  %%%%%%%%%%%%%%%%%%%%%%%%%%
\section{Introduction}
It is known that angular momentum of light has a spin part associated with polarization\cite{beth1936mechanical} and an orbital part associated with spatial distribution\cite{allen1992orbital}. 
Cylindrical vector beam (CVB)\cite{freund2001polarization,zhan2009cylindrical,wang2010new}, acting as a solution of vectorial Helmholtz equation\cite{hall1996vector,wang2007generation}, combines the two parts of angular momentum. 
Radial polarization and azimuthal polarization are the most conspicuous CVBs. 
Under tight focusing\cite{pu2010tight,porfirev2016polarization,otte2017tailored}, radial polarization possesses a sharper focal spot than a homogeneously polarized beam\cite{youngworth2000focusing,dorn2003sharper}, while azimuthal polarization can be focused into a hollow spot\cite{zhan2002focus}. 
These peculiar properties are useful for many applications such as particle manipulation
\cite{grier2003revolution,zhan2004trapping,man2018optical,turpin2013optical,weng2014creation}, microscopy\cite{hell1994breaking,biss2006dark,novotny2001longitudinal}, material processing \cite{kraus2010microdrilling,hamazaki2010optical,hnatovsky2013role}, near-field optics \cite{ciattoni2005azimuthally}, and nonlinear optics\cite{bouhelier2003near}. 
Recently, the degrees of freedom of CVB are extended by ray-like trajectories\cite{Shen:20}, so that CVB also presents growing potentials in the area of optical encoding\cite{milione2015using} and optical communications\cite{gibson2004free,bozinovic2013terabit,milione20154,jia2019mode}.

There are many methods to generate CVBs. 
Special intracavity resonator could
directly generate CVB from a laser when the cavity geometry is precisely controlled into a frequency-degenerate state\cite{kozawa2005generation,ahmed2007multilayer,naidoo2016controlled,wei2019generating,Wang:21}.
Meanwhile, single-element CVB generators have been introduced to tailor Pancharatnam-Berry phases \cite{marrucci2006optical,cardano2012polarization,mueller2017metasurface,devlin2017arbitrary}, supporting to modulate a couple of orthogonal polarization bases with conjugate phase distributions, 
to generate CVBs from a given basic laser mode. 
All these methods need to switch in different elements to generate CVBs with different topological charges, so that inconvenient in experiments that involve multiple CVBs.
Otherwise, implementations usually contain interferometers, where two orthogonal polarizations are modulated by spatially programmable devices individually and are then combined together\cite{rosales2017simultaneous}. 
Though ingenious devices with high 
robustness\cite{chen2011generation,chen2014generation,liu2019compact} are developed,
their complexity of modulating two individual parts still set them bulky and redundant. 
Therefore, it is significant to create a compact passive device to generate CVBs with arbitrary topological charges.

In this letter, we provide and demonstrate a passive device to generate CVBs with arbitrary azimuthal and radial topological charges. 
It realizes a polarization-selective $\pi$ Gouy phase mode converter\cite{beijersbergen1993astigmatic,jia2018integrated} via a couple of specially fabricated polarization-selective cylindrical lenses. 
The device could remain unchanged no matter the topological charges of CVBs. 
It even adapts to generate superimposed CVBs by employing the corresponding wavefront modulation of homogeneous polarization. 
The device creates a robust connection between homogeneous polarizations with transverse modes and CVBs, where homogeneous polarization can be marked on basic Poincar\'e sphere (PS) and CVBs can be marked on high-order Poincar\'e sphere (HOPS)\cite{milione2011higher}. 
Mapping between states on the basic PS and the first-order HOPS is experimentally proved by comparing Stokes parameters of five representative states with theoretical values. 
Its extension to generate states set on other HOPSs is verified by collecting petals-like intensities.

\section{Theory}
\begin{figure*}[htbp]
\centering
\includegraphics[width=\linewidth]{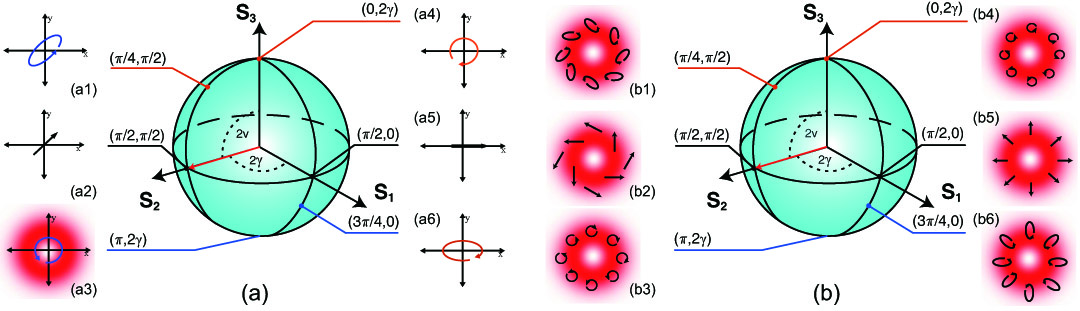}%\fbox{}deleted
\caption{(a) Basic Poinar\'e sphere. Selected points a$1\sim$ a$6 $ on the surface include polarization states from linear polarization to elliptical polarization. (b) The first-order HOPS. Selected points b$1\sim$ b$6 $ represent six CVBs on the surface. 
$2\upsilon$ is polar angle, $2\gamma$ is equator angle of arbitrary state vectors directing to points on the surfaces.  
Points a$1\sim$ a$6 $ map b$1\sim$ b$6$ one by one, which is just realized by the proposed device. }
\label{PoincareSphere}
\end{figure*}
Here Laguerre-Gaussian (LG) modes, a complete set for presenting transverse modes, is expressed with    
%For the sake of convenience in modes conversion, its 
complex amplitude 
\begin{equation}
{u_{\ell,p}}(r,\phi ,z) = {C_{\left|\ell\right|,p}(r,z)}\exp ( - i\ell\phi ),\label{upl}
\end{equation}
where $r$, $\phi$, and $z$ are spatial polar coordinates, $\ell$ and $p$ are indices of azimuthal and radial coordinates respectively. $\ell$ is an integer and $p$ is a natural number.  
$C_{\left|\ell\right|,p}$ marks a real coefficient related to $\left|\ell\right|$ and $p$, while 
$\exp ( - i\ell\phi )$ denotes spiral phase of wavefront which represents orbital angular momentum (OAM) of light with topological charge $\ell$. 
Employing Dirac notations, OAM state is expressed with $ \left|\ell\right\rangle$,
and polarization is characterized with circular bases $\lbrace\left|R\right\rangle,\left|L\right\rangle\rbrace$. 
They construct direct product state 
\begin{equation}
\begin{split}
\left| {\psi (2\upsilon ,2\gamma )} \right\rangle  &= \left(\cos{\upsilon}\left| {R} \right\rangle {{\rm{e}}^{-i\gamma}} + \sin  {\upsilon} \left| {L} \right\rangle {{\rm{e}}^{i\gamma}}\right)\otimes \left|\ell\right\rangle \\
&=\cos{\upsilon}\left| {\ell,R} \right\rangle {{\rm{e}}^{-i\gamma}} + \sin  {\upsilon} \left| {\ell,L} \right\rangle {{\rm{e}}^{i\gamma}},\label{SPS}
\end{split}
\end{equation}
which can be set on the surface of basic PS shown in Figure (\ref{PoincareSphere}a), %with spherical coordinates $2\gamma$ and $2\upsilon$, %the spherical coordinates of the polarization state, 
where $2\upsilon$ is polar angle ranging in the region $\left(0,\pi\right)$, $2\gamma$ is equator angle ranging in $\left(0,2\pi\right)$, %$\left| {\ell,R} \right\rangle \equiv \left| \ell \right\rangle \otimes \left| R \right\rangle$, 
and three spatial axes represent Stokes parameters $\rm{S}_1$, $\rm{S}_2$ and $\rm{S}_3$. %$\left| {\ell,L} \right\rangle \equiv \left| \ell \right\rangle \otimes \left| L \right\rangle$. 
CVB is defined with
\begin{equation}
\left| {\psi_\ell (2\upsilon ,2\gamma )} \right\rangle  = \cos{\upsilon}\left| {\ell,R} \right\rangle {{\rm{e}}^{-i\gamma}} + \sin  {\upsilon} \left| {-\ell,L} \right\rangle {{\rm{e}}^{i\gamma}}.\label{HOPS}
\end{equation}
which can be unified in an analytic model referring to HOPS\cite{milione2011higher}, where the simplest is first-order HOPS ($\ell=1$), as shown in Figure (\ref{PoincareSphere}b). 
The definition of spherical surface is similar with the circumstance of basic PS, except the bases change from $\lbrace\left|R\right\rangle,\left|L\right\rangle\rbrace$ to $\lbrace\left|\ell,R\right\rangle,\left|-\ell,L\right\rangle\rbrace$. 

Both radial polarization and azimuthal polarization are located on the equator of first-order HOPS. 
Denoting $\left|R\right\rangle = \left(\left|H\right\rangle-i\left|V\right\rangle\right)/\sqrt{2}$, $\left|L\right\rangle = \left(\left|H\right\rangle+i\left|V\right\rangle\right)/\sqrt{2}$, Eq. (\ref{HOPS}) indicates radial vector beam is marked with $\left| {\psi_1 (\pi/2,0 )} \right\rangle $, and azimuthal vector beam is notated with $\left| {\psi_1 (\pi/2 ,\pi )} \right\rangle $\cite{Tovar:98}. 
For convenience, they are abbreviated with spherical coordinates $(\pi/2,0 )$ and $(\pi/2 ,\pi )$. % in the form of $(2\upsilon ,2\gamma )$. 
The corresponding states with the same coordinates $(\pi/2,0 )$ and $(\pi/2 ,\pi )$ for basic PS are $\left|H\right\rangle$ and $\left|V\right\rangle$ polarizations respectively. %, which are also marked with $(\pi/2,0 )$ and $(\pi/2 ,\pi )$ respectively on PS. 
Figure (\ref{PoincareSphere}) elucidates the connection between basic PS and first-order HOPS via several representative points on the sphere. 
Points a$1\sim$ a$6 $ selected on the surface of basic PS shown in Figure (\ref{PoincareSphere}a) include polarization states evolving from horizontal polarization $(\pi/2,0 )$ to diagonal polarization $(\pi/2,\pi/2 )$ along with equator then turning to a general elliptic polarization $(\pi/4,\pi/2 )$ along with longitude line. 
Each points of a$1\sim$ a$6 $ correspond to points of b$1\sim$ b$6 $ on the same positions of first-order HOPS, which represents CVBs as shown in Figure (\ref{PoincareSphere}b).

\section{Implementation}
\begin{figure}[htbp]
\centering
\includegraphics[width=\linewidth]{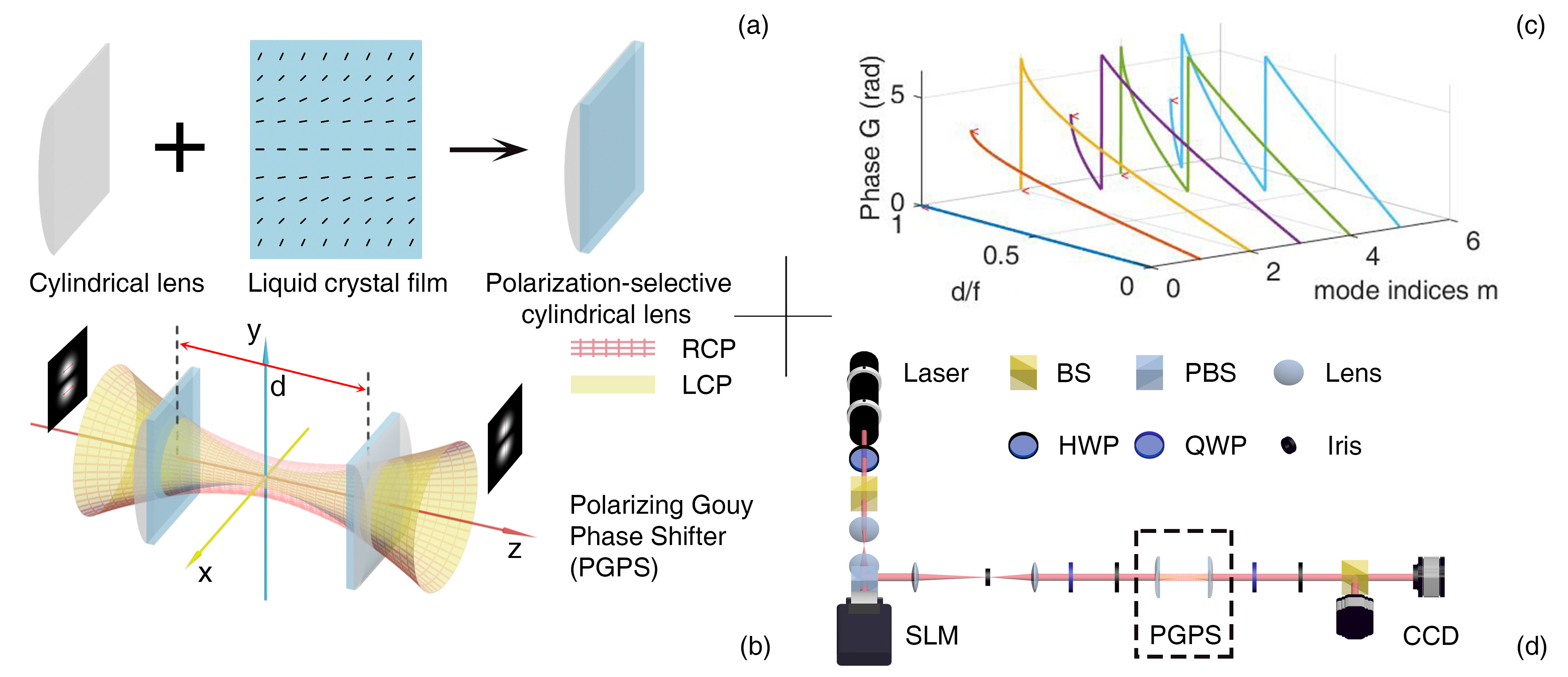}%\fbox{}deleted
\caption{(a) Configuration of polarization-selective cylindrical lens. (b) Polarizing Gouy phase shifter (PGPS), which contains two symmetric polarization-selective cylindrical lenses and operates on LCP only. (c) Gouy phases accumulated by PGPS for modes with different indices $m=0\sim 5$, the special points involved in the device are marked with red angles. (d) Schematic of the experimental setup for generating arbitrary CVBs. }
\label{Setup}
\end{figure}

Comparing Eq. (\ref{SPS}) with Eq. (\ref{HOPS}), it is found that $\left| {\ell,L} \right\rangle$ is converted to $\left| {-\ell,L} \right\rangle$ and $\left| {\ell,R} \right\rangle$ remains unchanged. Thus, the device needs to have two functions: response of polarization  and inversion of index $\ell$ .

Inspired by the design of Q-plate \cite{marrucci2006optical}, we employ liquid crystal films coated on a pair of cylindrical lenses 
to realize the conversion of polarization-selective response. 
As shown in Figure (\ref{Setup}a), polarization-selective cylindrical lens is constructed with isotropic glass lens and inhomogeneous liquid crystal film. 
Isotropic glass lens provides a common dynamic phase %in amount of Eq. (\ref{clens}) 
with thickness $y^2/{2f(n-n_0)}$ regardless of polarization, where $y$ is chosen as the converging direction of cylindrical lens, $n$ is the refractive index of the glass material, $n_0$ is the refractive index of air and $f$ is the designed focal length. 
The operator of common glass lens is expressed with $\exp{(+i\phi_1)}(\left|L\right\rangle\left\langle L \right|+\left|R\right\rangle\left\langle R \right|)$, where $\phi_1=-{ky^2}/{2f}$ according to Fresnel paraxial approximation. 
Liquid crystal film provides extra geometric phase\cite{simon1988evolving,kwiat1991observation,bomzon2001pancharatnam,bomzon2002space,oh2008achromatic} by metallic distribution of fast axis' angle $\theta={ky^2}/{4f}$ related to the direction of $x-$axis, written with operator $\exp (-i \phi_2)\left|L\right\rangle\left\langle R \right|+\exp (+i\phi_2)\left|R\right\rangle\left\langle L\right|$, in which $\phi_2=-{ky^2}/{2f}$. 
The compound operator, expressed with 
\begin{equation}
\exp [i(\phi_1-\phi_2)]\left|L\right\rangle\left\langle R \right|+\exp [i(\phi_1+\phi_2)]\left|R\right\rangle\left\langle L\right|=\left|L\right\rangle\left\langle R \right|+\exp (-i{ky^2}/{f})\left|R\right\rangle\left\langle L\right|,
\end{equation}
means %arbitrary phase distribution  can be accomplished in each polarization bases. 
 a polarization-selective cylindrical lens\cite{hasman2003polarization} with $y$ directional focal length $f/2$ operating on $\left|L\right\rangle$ is fabricated. 

The inversion of index $\ell$ derives from the coefficients conversion of Hermite-Gaussian (HG) modes.  
For complex amplitude of HG mode, $u_{nm}^{HG}$, $n$ and $m$ are two indices corresponding to $x$ and $y$ coordinates in the transverse plane. 
It is shown that a LG mode, %with complex amplitude 
$u_{nm}^{LG}$, can be decomposed into a set of HG modes with the same order $N$ ($N=n+m=2p+\left|\ell\right|$), written as $u_{nm}^{LG}(x,y,z) = \sum\limits_{k = 0}^N {{i^k}} b(n,m,k)u_{N - k,k}^{HG}(x,y,z)$. 
Real coefficient $b(n,m,k)$ is given by
\begin{equation}
b(n,m,k) = {\left( {\frac{{(N - k)!k!}}{{{2^N}n!m!}}} \right)^{{1 \mathord{\left/
 {\vphantom {1 2}} \right. 
 \kern-\nulldelimiterspace} 2}}}\frac{1}{{k!}}\frac{{{d^k}}}{{d{t^k}}}{[{(1 - t)^n}{(1 + t)^m}]_{t = 0}},
\label{bnmk}
\end{equation}
where $t$ is a continuous parameter around zero point. % and $k$ is defined in former summation. 
Exchanging $n$ and $m$ in Eq. (\ref{bnmk}), it is deduced that $b(m,n,k) = {( - 1)^k}b(n,m,k)$.
By definition $\ell=n-m$ in LG mode, extra factor $(-1)^k$ fits the conversion coefficients from $\ell$ to $-\ell$. 
Because polarization-selective cylindrical lenses are set in $y$ direction, astigmatic Gouy phase $G_y$\cite{beijersbergen1993astigmatic} is used to supply such an extra factor associated with mode index $m$, or written as $\exp(-imG_y)$. 

As shown in Figure (\ref{Setup}b), two polarization-selective cylindrical lenses are set on the symmetrical positions relative to original point $\rm O$. 
The condition for first piece of lens is $\phi_1-\phi_2=0$, causing left-handed circular polarization (LCP) becomes right-handed circular polarization (RCP) accompanied with converging effect while RCP becomes LCP without any side effect. 
Correspondingly, the second piece of lens, which is set in symmetrical position (flipped), turns RCP back to LCP with converging effect. 
Such that, only LCP of incident light accumulates $G_y$\cite{feng2001physical} between the two polarization-selective cylindrical lenses. 
The amount of $G_y$ is decided by the distance between two lenses, $d$, and designed focal length of them, $f/2$, under proper coupling conditions. 
Exactly, $ G_y =2\arcsin   d/f$, where $0\textless d\le f$. 
Phase of wavefront performs a $2\pi$ period, so the mode-dependent phase takes effects along the instruction of $G=\mod(m\cdot G_y,2\pi)$. 
Figure (\ref{Setup}c) shows amounts of $G$ for $m=0\sim 5$ in sequence. 
When $d/f=1$, special points marked with red angles performs like a phase switch between $0$ and $\pi$ along with $m$ axis, showing factor $(-1)^m$ is attained and index $\ell$ becomes $-\ell$ successfully. 
$d/f=1$ means $G_y=\pi$, so the device is called $\pi$ phase polarizing Gouy phase shifter ($\pi-$PGPS).

\begin{figure}[htbp]
\centering
\includegraphics[width=\linewidth]{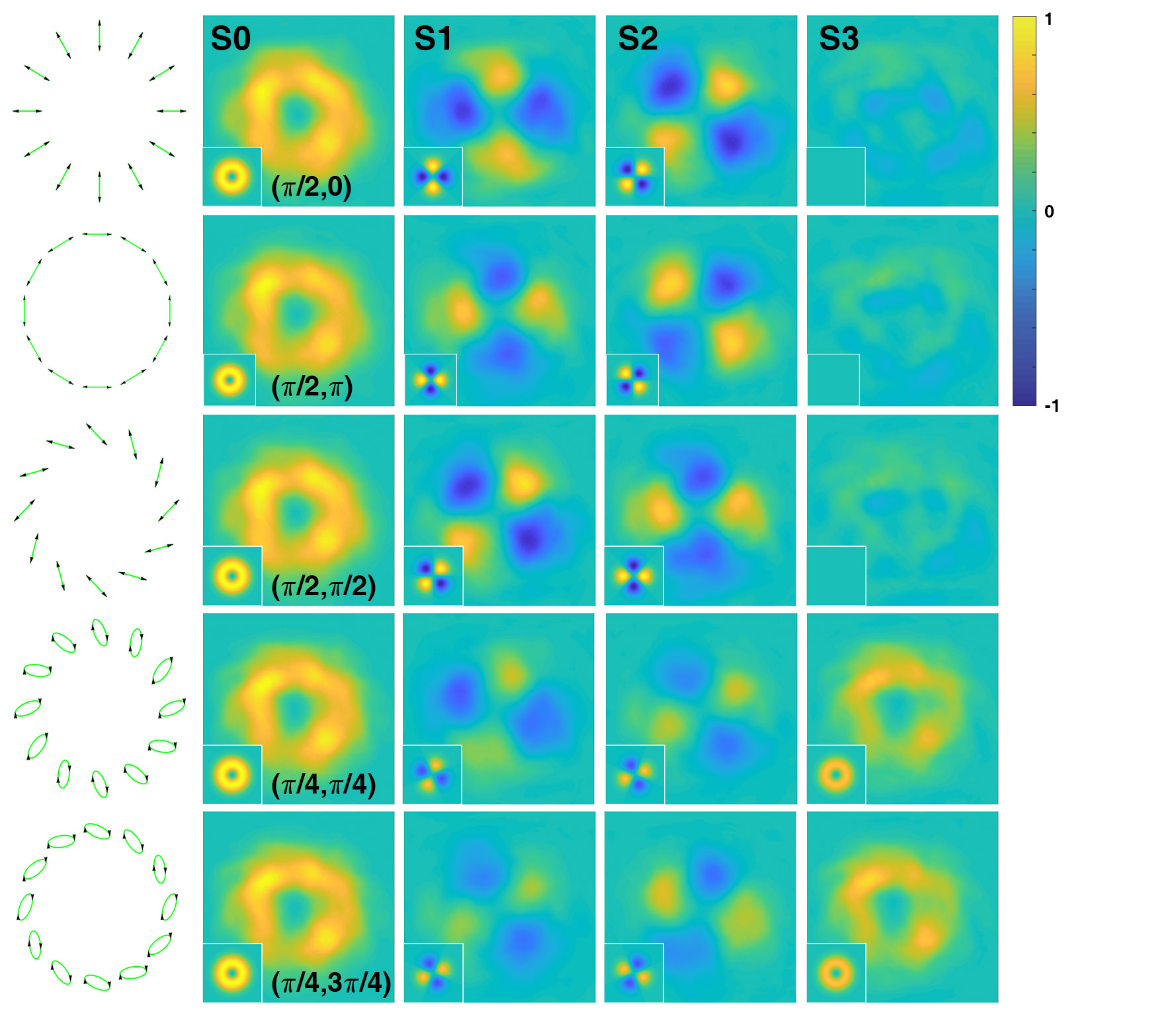}%\fbox{}deleted 
\caption{Stokes parameters of CVBs. The left column displays the tailoring polarization vectors of CVBs, followed by columns of Stokes parameters $S0\sim S3$. Subpictures in lower-left conner are theoretical values. CVBs are sampled from the first-order HOPS, where $\left(\pi/2,0\right)$, $\left(\pi/2,\pi\right)$, $\left(\pi/2,\pi/2\right)$, $\left(\pi/4,\pi/4\right)$ and $\left(\pi/4,3\pi/4\right)$ represent their positions on spherical surface. } 
\label{lattices} 
\end{figure} 
Figure (\ref{Setup}d) shows the experimental scheme. He-Ne laser derives a $632.8$ nm  Gaussian beam whose polarization is projected to horizontal by a half wave plate (HWP) and a polarizing beam splitter (PBS). 
Two lenses constitute an expander to provide an almost plane wave for the spatial light modulator (SLM, Holoeye, Pluto-VIS-016). 
Fork-like holograms\cite{bolduc2013exact} are loaded on the screen of SLM to produce LG modes. 
Beam splitter (BS) ensures light beam propagates in correct path. 
Two lenses and Iris select first order of diffracted light after SLM with reducing them into an appropriate scale. 
The characteristics of SLM determine that the original polarization of the selected beam is horizontal. 
If a pre-production of polarization is necessary, a quarter wave plate (QWP) and a HWP will be included in the optical circuit. 
Then the $\pi-$PGPS takes effects to generate %corresponding 
CVBs. 
Following QWP, HWP, PBS and a charge coupled device (CCD) constitute an examination framework of Stokes parameters. 
By changing angles of the fast axes of QWP and HWP relative to $x$ coordinate, polarizations can be reconstructed with intensities recorded by CCD. 
Exactly, Stokes parameters can be computed via\cite{schaefer2007measuring,wang2018optically}
\begin{subequations}
\begin{align}
{S_0} &= I\left( {{0^\circ },{0^\circ }} \right) + I\left( {{0^\circ },{{45}^\circ }} \right),\\
{S_1} &= I\left( {{0^\circ },{0^\circ }} \right) - I\left( {{0^\circ },{{45}^\circ }} \right),\\
{S_2} &= I\left( {{{45}^\circ },{{22.5}^\circ }} \right) - I\left( {{{45}^\circ },{{67.5}^\circ }} \right),\\
{S_3} &= I\left( { {{45}^\circ },{0^\circ }} \right) - I\left( {-{{45}^\circ },{0^\circ }} \right),
\end{align}\label{Stocks}
\end{subequations}
where $I$ represents collected intensities, the first angle in the bracket is of QWP, the second angle is of HWP. 
Results for S0$\sim$ S3 are revealed in Figure (\ref{lattices}). 
The quality can be evaluated by comparing the outcomes computed from collected intensities with corresponding values (lower-left conner) computed from simulated intensities.  
Radial polarization $\left(\pi/2,0\right)$ and azimuthal polarization $\left(\pi/2,\pi\right)$ are shown in the first two lines of Figure (\ref{lattices}). 
For common instances, states marked with $\left(\pi/2,\pi/2\right)$, $\left(\pi/4,\pi/4\right)$ and $\left(\pi/4,3\pi/4\right)$ on the first-order HOPS are presented in the last three lines. 
All the experimental results agree well with the simulated values, so that the device is effective in generating CVBs on the surface of the first-order HOPS. 
Obviously, all elements in the installation take effects in the reverse propagation, thus proving the device constructs a credible and reversible connection between basic PS and first-order HOPS successfully. 

\begin{figure}[htbp]
\centering
\includegraphics[width=\linewidth]{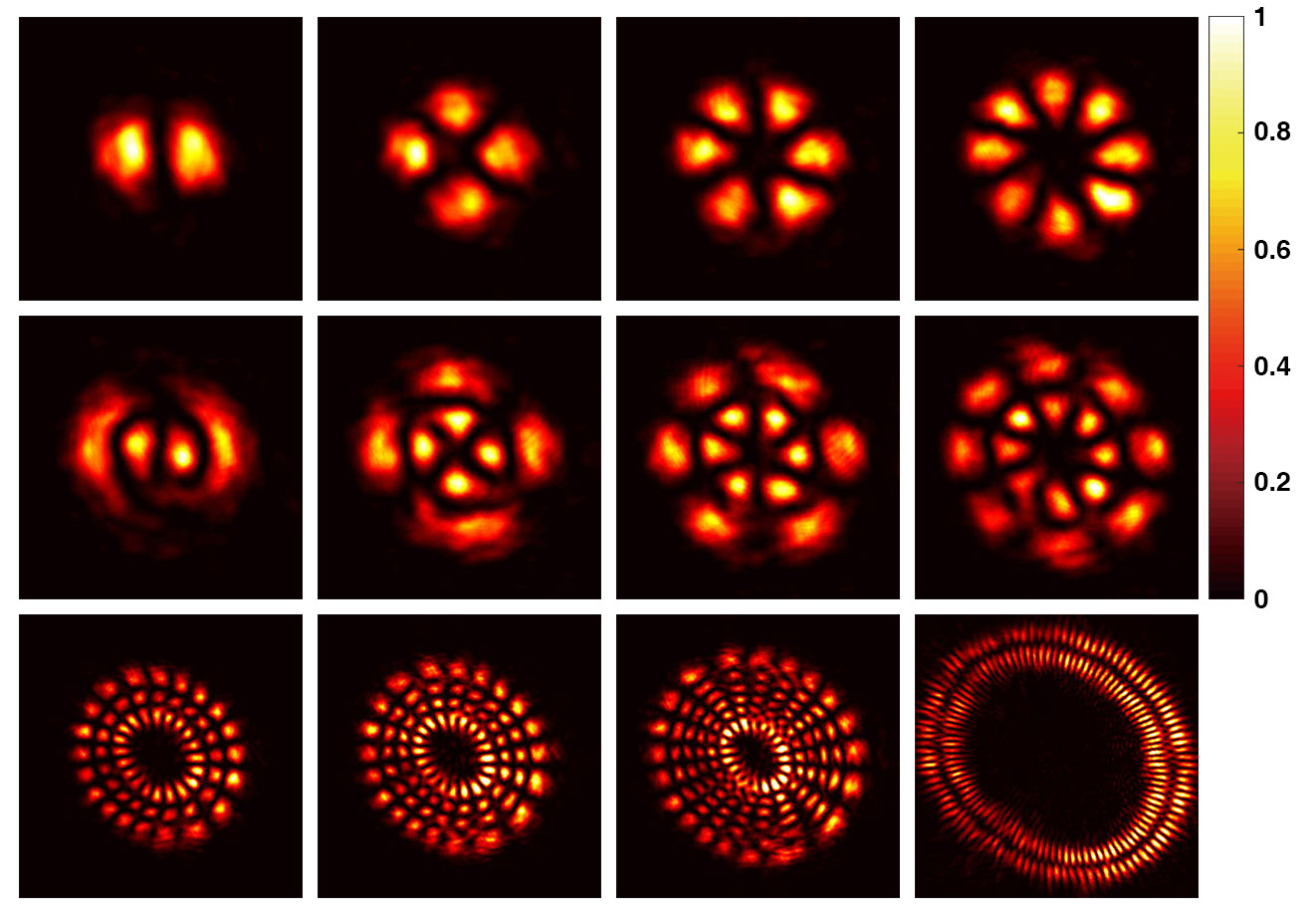}
\caption{Petals-like profiles collected by CCD when lunching high order LG modes.
The first row is set for $p=0$, $\ell=1,2,3,4$, and the second row is for $p=1$, $\ell=1,2,3,4$ respectively. 
The third row exhibits results for $\ell=10$, $p=2,3,5$ and an ultra-high order $\ell=50$, $p=1$. }
\label{HigherPedals}
\end{figure} 
As for $\left|\ell\right| \textgreater 1$, higher-order HOPSs are constructed. Representatively, a state on the equator of HOPS can be examined by casting it into horizontal polarization via transmitted port of PBS. 
The transmitted intensity performs special distribution which satisfies the petals-like shape, denoted by $I = \left|\left\langle {H} \mathrel{\left | {\vphantom {H { \psi_{\ell}(\gamma,\upsilon) }}} \right. \kern-\nulldelimiterspace} { \psi_{\ell}(\gamma,\upsilon) } \right\rangle_p \right|^2$, where subscript $p$ is the same with the index of LG mode. 
Combined with Eqs. (\ref{upl}) and (\ref{HOPS}), it is calculated that there are $2\left|\ell\right|$ pieces of intensity petals, so the number of petals can be exploited to characterize the azimuthal index $\ell$ of CVB, or called the order of HOPS. 
Eq. (\ref{upl}) shows that radial index $p$ is separated from the operation of $\ell$, meaning subscript $p$ is an individual parameter in the construction of HOPS. 
Real coefficient $C_{\left|\ell\right|,p}$ remains unchanged, and azimuthal phase $\exp ( - i\ell\phi )$ is reversed to $\exp ( i\ell\phi )$ under the conversion of the proposed device, so that each LG mode corresponds to a unique (non-degenerated) CVB. 
Naturally, complete LG modes and CVB modes are both defined with two topological charges $\ell$ and $p$ as their indices mapping with the same order of each other. 
Expression of CVB containing $p$ is written as
\begin{equation}
\left| \psi_{\ell}(2\upsilon,2\gamma) \right\rangle_p  = \cos{\upsilon}\left| {\ell,R} \right\rangle_{p} {\rm e}^{-i\gamma} + \sin{\upsilon}\left| { - \ell,L} \right\rangle_{p}{\rm e}^{i\gamma},\label{VVBform_p}
\end{equation}
where $\left| \psi_{\ell}(2\upsilon,2\gamma) \right\rangle_p \equiv  \left| \psi_{\ell}(2\upsilon,2\gamma) \right\rangle \otimes \left|p\right\rangle$. 
The general function of the device is illustrated by 
\begin{equation}
\mathbb{G} \left| {\psi (2\upsilon ,2\gamma )} \right\rangle_p = \left| {\psi_\ell (2\upsilon ,2\gamma )} \right\rangle_p,
\end{equation}
containing the generation of CVB with both $\ell$ and $p$ indices. 
$\mathbb{G}$ marks the operator of our device. 
In experiment, LG modes with $\ell=1,2,3,4$ and $p=0,1$ are generated by SLM. 
As shown in Figure (\ref{Setup}d), removing the two HWPs and two QWPs around PGPS from the circuit, petals-like intensities are collected directly after PGPS and a PBS. 
In Figure (\ref{HigherPedals}), the first row is set for results of $p=0$, the second row is for $p=1$. 
These pictures represent that the fabricated device is effective for CVB with both $\ell$ and $p$ indices. 
States on other ultra-high order HOPS are tested by lunching $\rm{LG}_{10,2}$ ($\rm{LG}_{\ell,p}$), $\rm{LG}_{10,3}$, $\rm{LG}_{10,5}$ and $\rm{LG}_{50,1}$ modes with results shown in the third line of Figure (\ref{HigherPedals}). 
In the collected intensities, outline of petals is always clear, showing the device works well with these higher-order CVBs.

\section{conclusion}
We propose and demonstrate a passive device based on $\pi-$PGPS to generate arbitrary CVBs with both $\ell$ and $p$ indices even in ultra-high order. 
The device simplifies existed schemes for generation and builds a solid connection between simple scalar field on basic PS and sophisticated CVBs on HOPS. 
Extensively, states on hybrid order Poincar\'e sphere (HyOPS)\cite{yi2015hybrid,liu2017generation} can be implemented with the help of Q-plate which provides a shift of $\ell$ index of CVB\cite{jia2019mode}. 
The device supplements a convenient operation for quantum information and communications experiment\cite{wang2012terabit,zhou2019using,wang2020high}, taking effects on the polarization-selective mode index inversion. 
Generally, it can be extended to arbitrary phase converter besides $G_y=\pi$ and support more splendid mode-dependent conversion of polarizations. 
The method is flexible to other techniques such as metamaterials
\cite{lin2014dielectric,deng2020malus,hu2019coherent} and metalens\cite{aieta2012aberration,khorasaninejad2016metalenses} 
which may help to miniaturize the optical device on chips\cite{piggott2015inverse,chen2018chip,wei2020high}.

\section*{Funding}
This work was in part supported by the National Natural Science Foundation of China (Grant Nos. 91736104, 11534008, 11974345 and 11804271).

%%%%%%%%%% If using BibTeX:
%\bibliography{sample}

%%%%%%%%%% If preparing manually:
% \newcommand{\enquote}[1]{``#1''}

% \bibitem{Zhang:14}
% Y.~Zhang, S.~Qiao, L.~Sun, Q.~W. Shi, W.~Huang, L.~Li, and Z.~Yang,
%   \enquote{Photoinduced active terahertz metamaterials with nanostructured
%   vanadium dioxide film deposited by sol-gel method,}
%   {\protect\JournalTitle{Optics Express}} \textbf{22}, 11070--11078 (2014).

% {Optical Society}, \enquote{{OSA Publishing},}
%   \url{http://www.osapublishing.org}.

% \bibitem{FORSTER2007}
% P.~Forster, V.~Ramaswamy, P.~Artaxo, T.~Bernsten, R.~Betts, D.~Fahey,
%   J.~Haywood, J.~Lean, D.~Lowe, G.~Myhre, J.~Nganga, R.~Prinn, G.~Raga,
%   M.~Schulz, and R.~V. Dorland, \enquote{Changes in atmospheric consituents and
%   in radiative forcing,} in \enquote{Climate Change 2007: The Physical Science
%   Basis. Contribution of Working Group 1 to the Fourth assesment report of
%   Intergovernmental Panel on Climate Change,}  S.~Solomon, D.~Qin, M.~Manning,
%   Z.~Chen, M.~Marquis, K.~B. Averyt, M.~Tignor, and H.~L. Miler, eds.
%   (Cambridge University Press, 2007).

% \end{thebibliography}

\end{document}